\documentclass[prb,twocolumn,floatfix]{revtex4}

\usepackage{graphics}
\usepackage{graphicx}
\usepackage{amsmath}
\usepackage[T1]{fontenc}

\def  \bsig    {\mbox{\boldmath$\sigma$}}

\begin{document}

\title{Anomalous Hall Effect due to the spin chirality in the Kagom\'{e} lattice}
\author{M. Taillefumier$^{1,2}$, B. Canals$^{1}$, C. Lacroix$^1$, V. K. Dugaev$^{1,2,3}$,
 and P. Bruno$^2$}
\affiliation{$^{1}$ Laboratoire Louis N\'{e}el - CNRS, 25 avenue des martyrs, BP 166,
38042 Grenoble Cedex 09, France\\
$^{2}$Max-Planck-Institut f\"ur Mikrostrukturphysik, Weinberg 2, D-06120 Halle, Germany\\
$^3$Department of Physics and CFIF, Instituto Superior T\'ecnico, Av. Rovisco Pais,
1049-001 Lisbon, Portugal}

\begin{abstract}
  We consider a model for a two dimensional electron gas moving on a
  kagom\'e lattice and locally coupled to a chiral magnetic texture.
  We show that the transverse conductivity $\sigma_{xy}$ does not
  vanish even if spin-orbit coupling is not present and it may exhibit
  unusual behavior. Model parameters are the chirality, the number of
  conduction electrons and the amplitude of the local coupling. Upon
  varying these parameters, a topological transition characterized by
  change of the band Chern numbers occur. As a consequence,
  $\sigma_{xy}$ can be quantized, proportional to the chirality or
  have a non monotonic behavior upon varying these parameters.
\end{abstract}

\maketitle

\section{Introduction}

In ferromagnetic systems, there are two contributions to the
transverse resistivity $\rho _{xy\text{ }}$: one is due to the usual
Lorentz force acting on the electrons when a magnetic field is
applied, $R_{0}B$, the second one, $R_{s}M$, is proportional to the
magnetization of the ferromagnet. This is called the anomalous Hall
effect.

The origin of this anomalous Hall effect has long been controversial
and both extrinsic (impurities) or intrinsic mechanisms were
discussed. Karplus and Luttinger\cite{Karplus1954} proposed that this
effect is a consequence of spin-orbit interaction in metallic
ferromagnets. Then it was argued that impurities give the main
contribution to the anomalous Hall effect and usually two mechanisms,
both due to spin-orbit coupling, contribute: one is known as side-jump
mechanism\cite{Berger1970,Berger1972} and it predicts that the Hall
resistivity \ $R_{s\text{ }}$ is proportional to the longitudinal
resistivity $\rho $. The second one\cite{Smit1954,Smit1958} is the
skew-scattering mechanism, which gives a contribution proportional to
$\rho ^{2}$.

In the recent years several groups measured anomalous Hall effect in
various systems which cannot be attributed to usual mechanisms
(skew-scattering or side-jump).\ A new intrinsic mechanism, related to
non-collinear spin configuration, with a ferromagnetic component, was
first proposed for manganites\cite{Matl1998,Ye1999,Chun2000}: in these
systems deviation from collinearity is a consequence of the
competition between double exchange, superexchange and spin-orbit
interactions. Then it was proposed that a similar mechanism works in
spin glass systems where spin configuration is highly non
coplanar\cite{Tatara2002,Kawamura2003}: in the weak coupling limit it
was shown that the Hall resistivity is proportional to the uniform
chirality parameter with a sign which depends on the details of the
band structure.\ This chirality parameter was defined by Tatara and
Kawamura\cite{Tatara2002} for three spins at sites i, j, k as a scalar
quantity: $%
\chi =\langle \mathbf{S}_{i}.(\mathbf{S}_{j}\times
\mathbf{S}_{k})\rangle $, being non-zero if the spins are
non-coplanar. Very recently several groups confirmed the existence of
such a contribution in typical AuFe spin
glasses\cite{Pureur2004,Taniguchi2004}.

In non-disordered systems, the same mechanism can work as soon as the
ordered magnetic structure is non-coplanar. This is realized in
several pyrochlore compounds such as Nd$_{2}$Mo$_{2}$O$_{7}$ for which
Taguchi et al\cite{Taguchi2001} proposed that an anomalous contribution
to the Hall effect is related to the umbrella structure of both Nd and
Mo moments in the long range ordered phase. Detailed studies of this
system and other pyrochlores (Sm$_{x}$Y$_{1-x}$Mo$_{2}$O$_{7}$,
Nd$_{x}$Y$_{1-x}$Mo$_{2}$O$_{7}$) have been
performed\cite{Taguchi2001,Taguchi2003,Yasui2003} leading to some
controversy: using neutron diffraction experiments in magnetic field
it is possible to calculate the T and H dependence of both Mo and Nd
chiral order parameter and in Ref.~\onlinecite{Yasui2003} it was
concluded that the chiral mechanism alone cannot explain the T-dependence of
the Hall coefficient in pyrochlores.

From the theoretical point of view, this mechanism is often called
''Berry phase contribution'' because a non-vanishing spin chirality is
associated with a non-vanishing Berry phase for conduction electrons
coupled via local Hund's exchange interaction to the spins.\ Ohgushi
et al\cite{Ohgushi2000} calculated the Hall effect in a Kagom\'{e}
lattice with a non-coplanar long range spin structure.\ They have
shown that, in the adiabatic limit, when the electron conduction spins
are aligned with localized spins at each site of the lattice (this
corresponds to infinite Hund's coupling), the Berry phase contribution
to Hall conductivity is quantized for some values of the band filling.
In this paper, we study a model which extrapolates between this strong
coupling limit and the weak coupling case studied by Tatara and
Kawamura\cite{Tatara2002}. We show also that the Berry phase
contribution does not depend only on the chirality, but also on the
strength of the local Hund's coupling and on the band filling.

\bigskip

\section{Model}
\subsection{Hamiltonian}

The aim of this work is to show how the transport properties of
electrons are influenced by two different contributions. First, the
electrons are restricted to move on a lattice and are therefore
experiencing its geometry. This results in a peculiar band structure.
Second, each electron has its spin locally coupled to a given
distribution of magnetic moments on each site of the lattice.

Both effects are taken into account in the following Hamiltonian:
\begin{equation}
\mathcal{H}=\sum_{\langle i,j\rangle ,\sigma }t_{ij}\left( c_{i\sigma
}^{\dagger }c_{j\sigma }+h.c.\right) -J\sum_{i}c_{i\alpha }^{\dagger }\left(
\mathbf{\sigma }_{\alpha \beta }\cdot \mathbf{S}_{i}\right) c_{i\beta }.
\label{eq:1}
\end{equation}
The first term describes the electrons moving on the lattice: $t_{ij}$
is the hopping integral between two neighboring sites $i$ and $j$;
$c_{i\sigma }^{\dagger }$ and $c_{i\sigma }^{{}}$ are the creation and
annihilation operators of an electron with spin $\sigma $ on the site
$i$. The second part of the Hamiltonian couples the electron spin to a
local moment on each site.

\begin{figure}[h]
\centering \includegraphics[width=8cm]{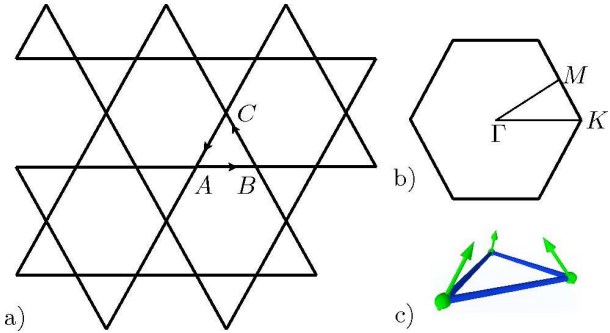}
\caption{(a)~The
Kagom\'e lattice is described as a triangular lattice of
triangles.
Its Bravais vectors are $a_1=(1,\, 0)$, $a_2=(-1/2 , \,
\protect\sqrt{3}/2)$, and $a_3=-(1/2 , \, \protect\sqrt{3}/2)$,
connecting respectively, the points $A$ to $B$, $B$ to $C$, and
$C$ to $A$.
(b)~The first Brillouin zone is an hexagon with the corners located at
$k=\pm (2\protect\pi/3)\, a_1$, $k=\pm (2\protect\pi/3)\, a_2$ and
$k=\pm (2\protect\pi/3)\, a_3$.
(c)~The umbrella structure on the triangular cell of the
kagom\'e lattice.} \label{fig:1}
\end{figure}

The coupling constant to each local moment ${\bf S}_i$ is $J$, and
these moments are treated below as classical variables. $\bsig
_{\alpha \beta }$ are the Pauli matrices. The underlying lattice is
the Kagom\'{e} lattice, depicted in Fig.~\ref{fig:1}. It is a two
dimensional tiling of corner sharing triangles, described as a
triangular lattice of triangles throughout this article.

This Hamiltonian has already been discussed in the limit of $J\to
\infty $ by Ohgushi et al.\cite{Ohgushi2000} In this limit the two
$\sigma =\uparrow ,\downarrow $ bands are infinitely splitted and
the model describes a fully polarized electron gas subject to a
modulation of a fictitious magnetic field, corresponding to the
molecular field associated with the magnetic texture.
The comparison with the present work will be done later on.
In a general case of finite $J$, the calculations must be done
numerically.
It is worth noting that the sign of $J$ is unimportant in this
classical treatment since changing $J$ to $-J$
is equivalent to an exchange of $\uparrow$ and $\downarrow$ spin
states.

\subsection{Parameters and motivation}

We consider the spin texture as a 'parameter' of the model.
The local moments ${\bf S}_i$ are described classically, and the
chosen magnetic phase is periodic, thus allowing to work in the
reciprocal space.
The choice of the magnetic arrangement is motivated
by two reasons.
First, we are interested in describing the anomalous transport
properties of the type already addressed in previous works
(see Refs.~[\onlinecite{Ohgushi2000,Taguchi2003,Onoda2003b}]) in which
it has been argued that the transverse conductivity may be related
to the chirality of the magnetic phase.
%
%
Second, an unusual mechanism of the AHE has recently been
proposed, based on an unconventional chiral magnetic ordering due
to geometrical
frustration\cite{Taguchi2003,Kageyama2001,Katsufuji2000,Yoshii2000} but
with controversial interpretations.
We therefore consider the present model as a simple model to
address these problems and clearly identify each relevant
contribution.

We consider the magnetic phase which has been obtained by studying
a pure spin model with the anisotropic Dzyaloshinskii-Moriya interactions
on the Kagom\'{e} lattice\cite{Elhajal2002}.
It consists of an umbrella of three spins per unit cell of the
Kagom\'{e} lattice (see Fig.~\ref{fig:1}).
Each umbrella can be described by the spherical coordinates of the
three spins $(\pi /6,\, \theta )$,
$(5\pi /6, \, \theta )$ and $(-\pi /2,\, \theta )$.
The angle $\theta $ ranges from $0$ (all the spins are perpendicular
to plane, and the corresponding ordering is ferromagnetic) to $\pi $.
It is worth noting that the usual three sublattice planar magnetic phase
belonging to the ground state manifold of the kagom\'e antiferromagnet,
and commonly denoted as the ``$q = 0$ phase'' corresponds to
$\theta = \pi / 2$.
%
%

In between, all phases are possible and chiral, where chirality is
defined as the mixed product of three spins on a plaquette
\begin{equation}
\chi _{ijk}={\bf S}_{i}\cdot ({\bf S}_{j}\times {\bf S}_{k})
=\frac{3\sqrt{3}}{2}\cos \theta \sin ^{2}\theta .
\label{eq:3}
\end{equation}%
All these phases ($\theta \in [ 0, \pi ]$) are therefore
translation-invariant but do not have time reversal symmetry.

We consider the local spins as classical spins with length M and
we introduce in eq.~\ref{eq:1} an effective coupling constant
$J_{0} = J M $ which allows to rewrite the hamiltonian as
\begin{equation}
\mathcal{H} =
    \sum_{\langle i,j\rangle ,\sigma }
     t_{ij}\left( c_{i\sigma}^{\dagger }c_{j\sigma }+h.c.\right)
- J_{0} \sum_{i}c_{i\alpha }^{\dagger }
    \left(\bsig_{\alpha \beta } \cdot
          \mathbf{n}_{i}\right) c_{i\beta }
\end{equation}
\noindent where $\mathbf{n}_{i}$ is a unit vector collinear to the
local moment ${\bf S_i}$.

Setting $ t=| t_{ij} |$
as the energy unit, our free parameters are the angle $%
\theta $, parametrizing the chirality of the magnetic texture, and
the value of $ J_{0}$ .
The Fermi level, through the filling factor $p$ of the bands,
can also be varied.
Each of these variables gives a different contribution as will be
discussed in the next sections.


\subsection{Physical quantities}
Before calculating any physical observables, we have to diagonalize
the Hamiltonian of Eq.~\ref{eq:1}. For this purpose, the Hamiltonian
is rewritten in the reciprocal space as
\begin{equation}
\mathcal{H}=\sum_{\mathbf{k}}\Psi _{\mathbf{k}}^{\dagger }h_{\mathbf{k}}\Psi
_{\mathbf{k}}\,+\,h.c.,
\end{equation}
with $\Psi _{\mathbf{k}}=(c_{A{\bf k}\uparrow },c_{B{\bf k}\uparrow
},c_{C{\bf k}\uparrow },c_{A {\bf k}\downarrow },c_{B {\bf k}\downarrow
},c_{C{\bf k}\downarrow })$, $A,B$ and $C$ are the corners of the
kagom\'e lattice unit cell (Fig.~\ref{fig:1}-(a)) and
\begin{equation}
c_{A{\bf k}\sigma}=\sum_j c_{A,j,\sigma} \,e^{i {\bf k}\cdot
  {\bf r}_{ij} },
\end{equation}
is the Fourier transform of $c_{A,j,\sigma}$ and $r_{Aj}=r_A + R_j$
with $R_j$ the lattice vector and $r_A$ the position of the moment
${\bf S}_A$ in the unit cell. $h_{\mathbf{k}}$ is a $6\times 6$ matrix
given by
\begin{widetext}
  \begin{equation}
    h_{\bf k}=\left(
      \begin{array}{cccccc}
        -J_{0} \cos\theta &p^1_k &p^3_k &-J_{0} \sin\theta e^{i \frac{\pi}{6}} &0 &0 \\
        p^1_k &-J_{0} \cos\theta &p^2_k &0 &-J_{0} \sin\theta e^ {i \frac{5\pi}{6}} &0 \\
        p^3_k &p^2_k &-J_{0} \cos\theta &0 &0 &i J_{0} \sin\theta \\
        -J_{0} \sin\theta e^{-i \frac{\pi}{6}} &0 &0 &J_{0} \cos\theta &p^1_k &p^3_k \\
        0 &-J_{0} \sin\theta e^{-i\frac{5\pi}{6}} &0 &p^1_k &J_{0} \cos\theta &p^2_k \\
        0 &0 &-i J_{0} \sin\theta &p^3_k &p^2_k &J_{0} \cos\theta\\
      \end{array}\right),
    \label{eq:6}
  \end{equation}
\end{widetext}with $t_{ij}=t$ and $p_{\mathbf{k}}^{i}=2t\cos (\mathbf{k}\cdot
\mathbf{a}_{i})$.

To calculate the Anomalous Hall Effect, we use the expression of
the off-diagonal conductivity using the following Kubo formula in
the limit of disorder-free electron gas\cite{Onoda2002}
\begin{eqnarray}
  \label{eq:7}
  \sigma_{xy}(\omega) &=&\frac{e^2\hbar}{S}\sum_{n\neq m}\sum_{ \mathbf{k}} \left(f_{n\mathbf{k}%
    }-f_{m\mathbf{k}}\right)\nonumber\\
  &\times&\frac{\langle n,\mathbf{k}|v_x|m,%
    \mathbf{k}\rangle \langle n,\mathbf{k}|v_y|m,\mathbf{k}\rangle} {%
    \left(\varepsilon_{n\mathbf{k}}-\varepsilon_{m\mathbf{k}}\right)
    \left(\varepsilon_{n\mathbf{k}}-\varepsilon_{m\mathbf{k}%
      }-\omega\right)},
\end{eqnarray}
where $S$ is the surface of the unit cell, $\varepsilon_{n\mathbf{k}}$
is the eigenvalue of matrix~(\ref{eq:6}) corresponding to the energy
of $n$th band for the wave vector $\mathbf{k}$, with the eigenvector
$|n,\mathbf{k}\rangle$. Here $v_i=\frac1\hbar\frac{\partial h_{%
    \mathbf{k}}}{\partial k_i}$ is the velocity of electrons
($i=x,y,z$). Taking the static limit of $\omega\rightarrow 0$,
Eq.~(\ref{eq:7}) becomes
\begin{eqnarray}
  \label{eq:8}
  \sigma_{xy}&=&\frac{e^2\hbar}{S} \sum_{n,\mathbf{k}}f_{n,\mathbf{k}} \sum_{m\neq n}\frac{%
    (v_x)^{\mathrm{nm}}(v_y)^{\mathrm{nm}}-(v_x)^{\mathrm{mn}}(v_y)^{\mathrm{mn}}%
  }{(\varepsilon_{n, \mathbf{k}}-\varepsilon_{m,\mathbf{k}})^2}  \notag \\
  &=& \frac{e^2}{\hbar S}\sum_{n,\mathbf{k}}f_{n,\mathbf{k}}\left[\mathbf{\nabla}_{\mathbf{k}%
    }\times \mathbf{\mathcal{A}}_{n,\mathbf{k}}\right]_z,
\end{eqnarray}
where $\mathbf{A}_{n\mathbf{k}} = -i \langle n \mathbf{k}|\nabla_{\mathbf{k}%
}| n \mathbf{k} \rangle$ is the geometric vector potential and $%
f_{n\mathbf{k}}$ is the Fermi-Dirac distribution function.

The expression (\ref{eq:8}) is similar to that obtained by Thouless
\textit{et al.}\cite{Thouless1982} in the context of the Quantized Hall
effect in two dimensions. In particular, when the band is completely
filled, the circulation of $\mathbf{A}_{n\mathbf{k}}$ or the flux of
Berry curvature defined as
$\mathbf{\Omega}_{n\mathbf{k}}=\mathbf{\nabla}_{\mathbf{k}}\times
\mathbf{A}_{n\mathbf{k}}$ over the first Brillouin zone is equal to
$2\pi i$ times an integer\cite{Thouless1982,Thouless1983} called the Chern
number $\nu_n$. Expression~(\ref{eq:8}) of the off-diagonal
conductivity finally becomes
\begin{eqnarray}
  \sigma_{xy} = \frac{e^2}{\hbar S} \sum_{n\mathbf{k}} f_{n\mathbf{k}}\mathbf{\Omega}%
  _{n\mathbf{k}} = \frac{e^2}{h} \sum_n\nu_n,  \label{eq:9}
\end{eqnarray}
where $\nu_n=\frac{1}{2\pi}\int f_{n\mathbf{k}} \Omega_{n\mathbf{k}}
d^2 {\bf k}$. As we shall see, this implies that the Hall conductivity
is quantized when the Fermi level lies in a gap.

The Chern numbers have the following properties: (i) the sum over
all bands is equal to $0$, (ii) if for some peculiar values of
parameters of the model, the energy bands $m$ and $n$ cross each
others at some point of the Brillouin zone, their respective Chern
number obey the following conservation rule
\begin{equation}  \label{eq:10}
  (\nu_m+\nu_n)_b=(\nu_m+\nu_n)_a
\end{equation}
where the indices $b,a$ refer to the values of the Chern number
before and after the bands are touching each other.\cite{Avron1983}
We will check each of these conservation rules in the next
sections.

It should be noticed that the sum over $\mathbf{k}$ in Eq.~(%
\ref{eq:9}) runs over all occupied states. This may seem unusual
as in  metallic systems we expect that only  the quasiparticles
near the Fermi surface contribute to transport properties.
However, it has been recently shown by Haldane\cite{Haldane2004}
that Eq.~(\ref{eq:9}) can be reconciled with this point of view
and reduced to  a sum over the states near the Fermi energy.

\section{Numerical results}

\label{sec:numerical-results}

This section is divided in two parts. In the first one, the band
structure is calculated for different angles $\theta$ and coupling
constants $J_0$. The evidences for two different regimes of couplings
are given, illustrated by the variation of Chern numbers. In the
second one, we focus on the transverse conductivity at zero
temperature. It is shown that the chirality may be a relevant quantity
for describing the transverse conductivity ($\sigma \propto \chi$) in
some ranges of coupling only for a half-filled band.

\subsection{Band structure and associated Chern numbers}

\begin{figure}[tbp]
  \centering \includegraphics[width=8cm]{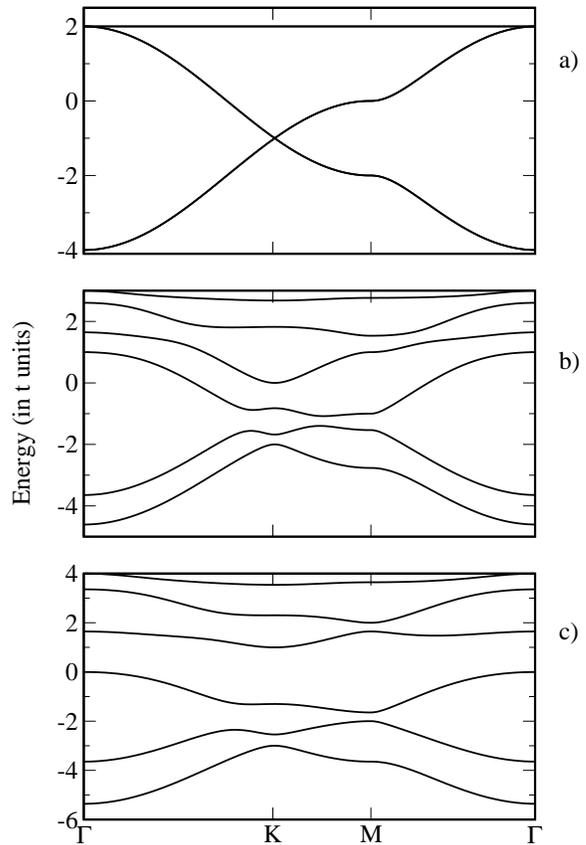} \caption{ (a) Energy
    spectrum of~(\ref{eq:5}) calculated for $J_{0}=0$. Each band is
    twice degenerate due to the spin degeneracy. (b) Energy spectrum
    calculated for $J_{0}=t$ and $\protect\theta=\protect\pi/3$. (c)
    Energy spectrum calculated for $J_{0}=2 t$ and $\protect\theta=%
    \protect\pi/3$. The critical value $J_c$ is equal to
    $4t/\protect\sqrt{7}%
    \approx 1.51\ t$}
\label{fig:2}
\end{figure}

The energy spectrum of the Kagom\'e lattice in the absence of
exchange interaction, is characterized by one flat band at $E=2t$
and two dispersive bands, which touch each other at the ${\bf
k}$-point $K$ of the Brillouin zone as shown in
Fig.~\ref{fig:2}.
This peculiarity has already been addressed in the context of spin
models as well as for the metallic systems\cite{Mielke1991}.

Once the coupling constant $J_{0}$ is nonzero, the energy spectrum
splits in two parts due to the spin dependant potential. For very
large values of $J_{0}$, the spectrum is divided into two groups of
three bands, which is the case studied by Ohgushi {\it et
  al.}\cite{Ohgushi2000}. In between, for nonzero but not too large
values of $J_{0}$, the spectrum can only be computed numerically,
except for the special cases of $\theta = 0$ or $\pi$, or for general
$\theta$ at high symetry points. For these
intermediate values of $J_{0}$, the splitting of the spectrum depends
qualitatively on two mechanisms.

First, the coupling $J_0$ gradually separates each group of three
bands taking them degenerate for $J_{0}=0$ to fully separated for
infinite $J_{0}$. Second, within each group of three bands, point
like degeneracies are lifted (see Fig.~\ref{fig:2}, points
$\Gamma$ and $K$) when switching on $J_{0}$, and finally restored
for $J_0 \to \infty$ (see Ref.~\onlinecite{Ohgushi2000}). The
numerical calculation of the spectrum shows that it is either
gapless for small values of $J_{0}$, or has gaps for higher
values.

\begin{figure}[h]
\centering \includegraphics[width=7cm,angle=-90]{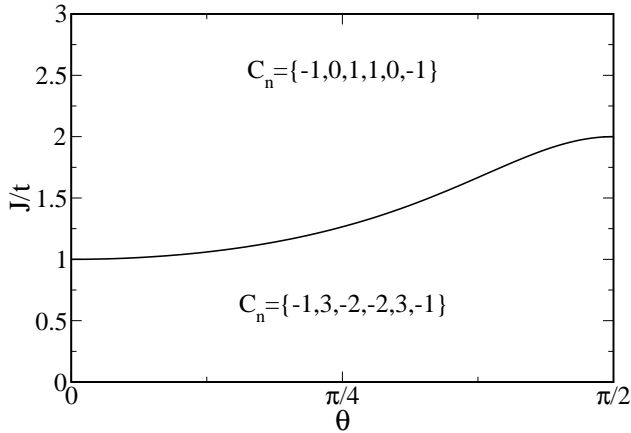}
\caption{The phase diagram in $J_{0}-\protect\theta$ variables.
The
  black line is given by Eq.~(\ref{eq:11}) and corresponds to the
  vanishing gap between the bands $2$ and $3$ (the gap between the
  bands $4$ and $5$ is also vanishing).}
\label{fig:3}
\end{figure}

When gaps open, it always occurs at the $M$ point of the Brillouin
zone. This allows to compute analytically the critical value of the
coupling as a function of chirality parameterized by the angle
$\theta$,
\begin{equation}  \label{eq:11}
J_c(\theta)=\pm\frac{2}{\sqrt{1+3\cos^2\theta}}.
\end{equation}
Using Eq.~(\ref{eq:11}) we can distinguish between two different
regimes (Fig.~\ref{fig:2}) depending on the value of $J_{0}$ as compare
to $J_c(\theta)$. These regimes are characterized by a particular
organization of the band structure as shown in Figs.~\ref {fig:2}-(b)
for $J_{0}=t$ and \ref{fig:2}-(c) for $J_{0}=2 t$ and different Chern
numbers. For $J_{0}>J_c(\theta)$, the Chern numbers associated to each
band are given by $-1,\, 0,\, 1,\, 1,\, 0,\, -1$ from the lower to the
upper band. Those numbers were obtained by Ohgushi \textit{et
  al.}\cite{Ohgushi2000} in the limit of $J_0\to \infty$. When $
J_{0}<J_c$, the Chern numbers associated to this configuration are
$-1,\, 3,\, -2,\, -2,\, 3,\, -1$ given in the same order as in the
previous case. The global sum rule $\sum_n \nu_n = 0$ is always
verified as well as the local one around the critical coupling where
two pairs of bands are crossing each other. It reads as
\begin{equation}
\nu_2 + \nu_3 = 1 \quad\mathrm{{and}\quad \nu_4 + \nu_5 = 1 }
\end{equation}
below and above the critical coupling $J_c$, respectively. The
corresponding phase diagram is shown on figure~\ref{fig:3}.

\subsection{Off-diagonal conductivity at $T=0$}

\begin{figure}[h]
  \centering
  \includegraphics[width=8cm]{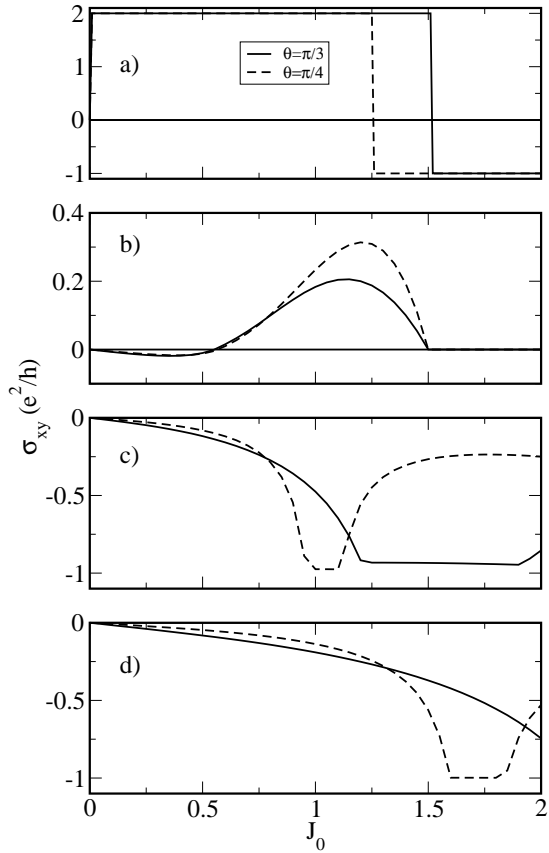}
  \caption{Off-diagonal conductivity $\sigma_{xy}(J)$ evaluated at
    $T=0$ for a filling factor $p=1/3$ (a), $p=1/2$ (b), $p=1/4$ (c)
    and $p=1/5$ (d) and $\theta=\pi/3$ (continuous lines) and
    $\theta=\pi/4$ (dashed lines).}
  \label{fig:5}
\end{figure}

The off-diagonal conductivity is computed using Eq.~(\ref{eq:7}) at
$T=0$ for different values of the chirality $\theta$ and the amplitude
of the exchange interaction $J_{0}$.
\begin{figure}[h]
\centering \includegraphics[width=8cm]{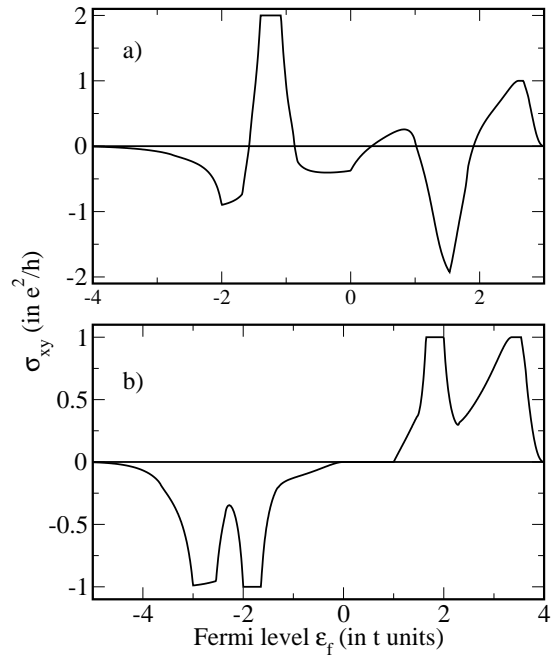} \caption{Off-diagonal
conductivity evaluated at $T=0$ as a function of
  $\varepsilon_f$ for $\protect\theta= \protect\pi/3$ and (a)
  $J_{0}=2\ t$ (b) $J_{0}=t$. In both cases, $\protect\sigma_{xy}$ is
  quantized when the Fermi level is lying in a gap. The values of the
  plateaus depend strongly on the topology of the energy spectrum.}
\label{fig:4}
\end{figure}
As shown in Fig.~\ref{fig:4} (the corresponding energy spectrum is
presented in Figure~\ref{fig:2},(b) and \ref{fig:2},(c)) at $T=0$,
the Hall conductivity is quantized when the Fermi level is lying
in the gap. The value of the plateaus depends explicitly on the
Chern number of the filled bands. When the Fermi level is in a
band, the sum~(\ref{eq:9}) can be reformulated in terms of a sum
over the Fermi level as shown by Haldane\cite{Haldane2004}
recently. One has to note that when the Fermi level and $\theta$
are fixed, any variation of $J_{0}$ around $J_c(\theta)$ may
induce a jump in the Hall conductivity. For instance with a
filling factor $p = 1/3$ (Fig.~\ref{fig:5}-a), the Fermi level
always lies in a gap but the off-diagonal conductivity jumps from
$-e^2/h$ for $J_{0}>J_c$ to $2e^2/h$ for $J_{0}<J_c$.

A similar jump can be obtained of $\theta$ is changed, while $J_0$ is
constant: such a change in $\theta$ can for example be induced by
application of an external magnetic field perpendicular to the
kagom\'e plane. In this case, large change of Hall conductivity (even
sign change) can be induced by magnetic field.

This is not the case for $p = 1/2$. For that filling factor, the
Hall conductivity varies smoothly when $J_{0}$ passes through the
point $J_{0}=3\,t/2$ which does not depend on $\theta$
(Fig.~\ref{fig:5}-b). When $J<3\,t/2$, the Fermi level crosses the
bands three and four resulting to a nonzero Hall conductivity.
When $J>3\,t/2$, the Fermi level is in the gap separating the band
three and four and the off-diagonal conductivity is given by the
sum off the Chern number of the first three bands which gives
zero.

Fig~\ref{fig:5} shows the variation of the Hall conductivity as a
function of $J_{0}$ for filling factors 1/4 (Fig.~\ref{fig:5}-c)
and 1/5 (Fig.~\ref{fig:5}-d): in both cases the Fermi level is not
in a gap and Hall conductivity does not exhibit any jump.

From these comments, it is clear that the chirality is never a
relevant parameter for the Hall conductivity when the Fermi level
lies in a gap, because $\protect\sigma_{xy}$ is then quantized and
obviously, not proportional to the chirality.

\begin{figure}
\centering
\includegraphics[scale=0.9]{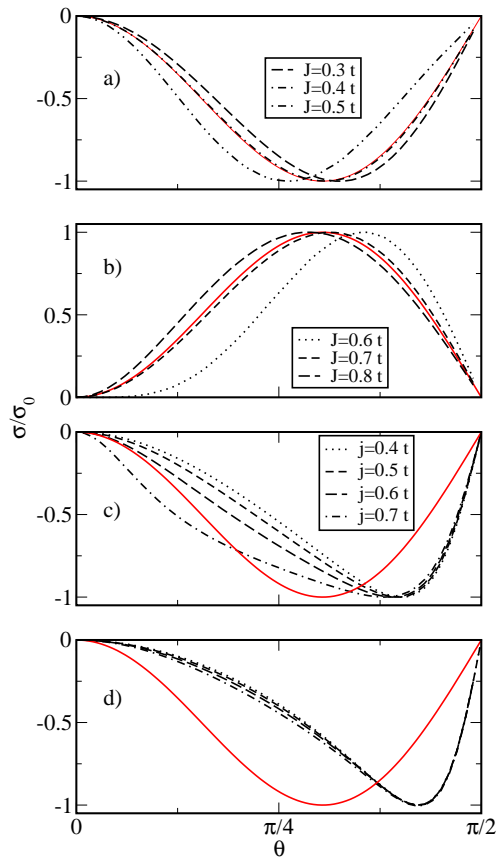}
\caption{$\protect\sigma_{xy}/\protect\sigma_0$ with
  $\protect\sigma_0=\max(|\protect\sigma_{xy}|)$ versus
  $\protect\theta$ angle, calculated for a filling factor $p = 1/2$
  (a)and (b) and $p=1/4$ (c) and $p=1/5$ (d) and different values of
  $J_{0}$. The continuous line represents the chirality given
  by~(\ref{eq:9}). The Hall conductivity is represented by the dashed
  lines for different values of the parameter $J_{0}$. We remark that
  $\protect\sigma_{xy}$ changes sign when $J_{0}$ tends to $0.6 t $.
  That change is related to the position of the Fermi level within the
  band but not associated to change of the Chern numbers. It is
  remarkable that the conductivity is more or less proportional to the
  chirality in a large range of couplings, up to $J_{0} \approx t$ as
  well as for large $\protect\chi$ values. For $p=1/4$ and $p=1/5$,
  one can show that the off-diagonal conductivity is not proportional
  to the chirality even for small values of $J_0$.}
\label{fig:6}
\end{figure}

Conversely, it is possible to choose a filling factor such that
the Fermi level is not within a gap. In Fig.~\ref{fig:6}, we
present $\sigma_{xy}$ versus $\theta$ for the filling factor $p =
1/2$ (a) and (b) and different values of $J_{0}$. For this filling
factor, the Fermi level is not in a gap if $J_{0}<3\,t/2$. Scaling
the conductivity by its maximal value in each case shows that it
is proportional to chirality in a large range of coupling $J_{0}$,
but also for large values of the chirality. This is somewhat
unexpected as the previous studies relating $\sigma_{xy}$ to
$\chi$ were restricted to small values of $\chi$\cite{Tatara2002}.
However, for different values of the band filling this is no
longer true as shown on figure~\ref{fig:6} for $p=1/4$
(\ref{fig:6}-c) and $p=1/5$ (\ref{fig:6}-d): for these filling
factors, $\sigma_{xy}$ is sensitive to the variation of the Chern
numbers when $J_{0}$ varies. In fact we found that the Hall
conductivity is proportional to chirality only for $p=1/2$ and
small enough $J_{0}$.

\section{Discussion}

The model studied in this paper cannot be directly applied to
materials, in which the Anomalous Hall Effect attributed to the
spin chirality has been observed.

However, some interesting features have been obtained in this model.
This work shows that transverse conductivity may be proportional to
chirality in a strong coupling regime, extending previous results
obtained in the weak coupling case\cite{Tatara2002} but this occurs only
for a half filled band. It is also shown that the dependence of
$\sigma_{xy}$ is not only determined by the chirality but also depends
strongly on the exchange coupling parameter and on the band filling
factor. $\sigma_{xy}$ can present a large variety of behaviors:
non-monotonic variation, sign change or plateaus can be observed.
Thus, even if the underlying magnetic phase chirality is the main
origin of this intrinsic anomalous Hall effect, it is far from being
entirely determined by the chirality.

We suggest that in real systems, variation of coupling J can be
induced by temperature: the temperature acts on the system in many
different ways. It acts on the magnetic configuration of the
localized spins $\mathbf{S}_i$ by changing the amplitude of the
magnetization, and on the position of the Fermi level in the band
through the Fermi-Dirac distribution.

We assume that the magnitude of each local moment
follows the mean field relation
\begin{equation}
M(T)= |{\bf S_i}|
    = M_{0}\sqrt{1-\frac{T}{T_{c}}}.  \label{eq:4}
\end{equation}%
where $T_c$ is the critical temperature. This can be described by
introducing in eq.~1 a temperature dependent coupling constant

\begin{equation}
J(T)=J_{0}\sqrt{1-\frac{T}{T_{c}}}.  \label{eq:5}
\end{equation}%
where $J_{0} = J M_0$ is the zero temperature exchange constant
and $J_{0}$ and $T_{c}$ are supposed to be independent.

\begin{figure}[h]
\centering
\includegraphics[scale=0.95]{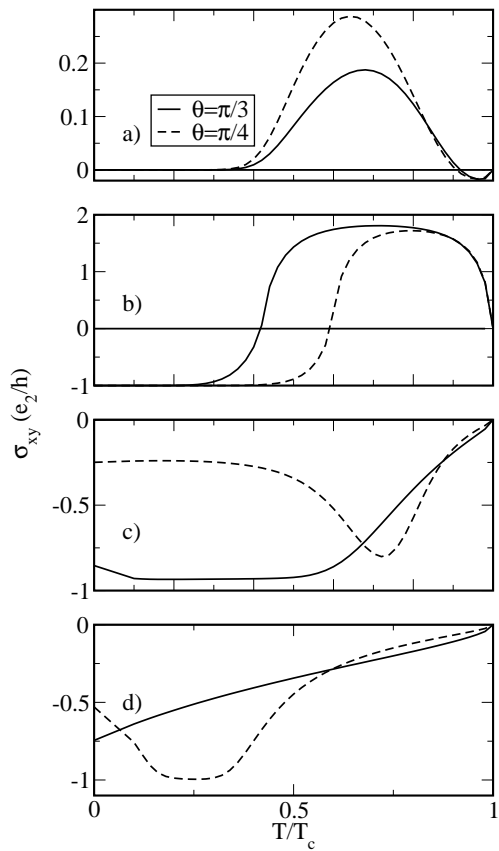}
\hspace{0.2mm}
\caption{Off-diagonal conductivity $\sigma_{xy}(T)$
for a filling
    factor $p=1/2$ (a), $p=1/3$ (b), $p=1/4$ (c) and $p=1/5$ (d) and
    $\theta=\pi/3$ (continuous lines) and $\theta=\pi/4$ (dashed
    lines).}
\label{fig:7}
\end{figure}

Consequently, decreasing the temperature from $T_{c}$ to $T=0$ is
equivalent to increase the exchange coupling from $J=0$ to $J=J_{0}$.
This increase of $J$ can be considered as done at zero temperature
since in a large range of parameters, the additional effect of
temperature (i.e. the $T$-dependance of the Fermi function in
Eq.~(8)), is significant only when the Fermi level is very close to a
band edge.

Qualitatively, this means that when temperature is decreased, one
moves vertically from the bottom to the top of the phase diagram in
Fig.~\ref{fig:3}: if $J_{0}$ is large enough, the temperature decrease
produces at some peculiar temperature $T_{\star }$ such that
$J(T_{\star })=J_{c}(T=0)$, a change in conductivity due to the change
of the Chern numbers. From this observation it follows that $\sigma
_{xy}$ can show abrupt changes with temperature. It is worth noting
that during that process, in the frame of our model, the magnetic
texture characterized by its $\theta $ angle has \textit{not} moved.
Of course, a change in $\theta $ can also be responsible for a sign
change.

To be more specific, we consider several examples presented in
Fig.~\ref{fig:7}.

We start with the filling factor $p = 1/2$, the critical temperature
$T_c = 0.1 t$ and $J(T=0)=2t$ (Fig.~\ref{fig:7}-a). With this choice of
parameters, the Fermi level lies in a gap when $J_{0}>3\,t/2$, which
means that at low temperatures (i.e., for large $J_{0}$), the system
is insulating. When the temperature is increased, $\sigma_{xy}$
remains equal to zero untill $T$ reaches the temperature $T_{1}$ for
which $J(T_{1})=3\,t/2$. Above this temperature, the gap closes,
making the transverse conductivity increasing continuously. It reaches
an extremum which is a function of the texture angle $\theta$. Just
below $T_c$, $\sigma_{xy}$ changes sign because of a subtle balance
between the states giving positive and negative contributions to the
conductivity, namely, the states from bands 3 and 4 (negative
contribution) and the states from band 2 (positive contribution).
These contributions may be globally explained by the associated Chern
numbers of these bands, at small $J_{0}$ : $\nu_3, \nu_4 = -2$ and
$\nu_2 = 3$.

For the filling factor $p = 1/3$ (Fig.~\ref{fig:7}-b), the behavior is
completely different. The transverse conductivity is quantized and
finite at zero temperature as the Fermi level is located in a gap.
When the temperature increases and reaches the value $T_{\star}$
defined by $J(T_\star)=J_c(\theta)$, for which gaps are finite but
sufficiently small to authorize the interband processes, $\sigma_{xy}$
continuously increases. Upon increasing temperature, the conductivity
increases and then changes its sign. This change of sign is not of the
same origin as we considered for $p = 1/2$. In the first case, the
Chern numbers are fixed but the temperature changes the balance
between the weights of each band. In the second example, bands 3 and 4
get crossed thus changing their Chern numbers. The variation of Chern
numbers produces the sign change in the transverse conductivity. Due
to thermal fluctuations, the extremum of conductivity does not
reach its quantized value of 2 (in units of $e^2/h$) but saturates at
1.8. The conductivity finally decreases down to zero when $T$ reaches
$T_c$, as it should be.

It is also possible to choose the filling factors that place the Fermi
level at zero temperature within a band. In these cases, there is no
generic behavior, and the angle of the spin texture plays an important
role. Some results are shown on figure~\ref{fig:7} for the different
filling factors $p = 1/4$ (c) and $p = 1/5$ (d).

Thus, these results show that in the mean field approximation a large
variety of behaviors are possible in our model. The next step would be
to study a model closer to the experimental situation of pyrochlores,
where the variation of magnetic structure with temperature and applied
field has been studied by neutrons experiments\cite{Yasui2003},
allowing to take into account the real cristallographic and magnetic
structures of these systems.

\begin{acknowledgements}
This work is partly supported by Université Joseph Fourier
(Grenoble), by FCT Grant POCTI/FIS/ 58746/2004 (Portugal) and by
Polish State Committee for Scientific Research under Grants
PBZ/KBN/ 044/P03/2001 and 2~P03B~053~25. V.D. thanks the Calouste
Gulbenkian Foundation in Portugal for support.
\end{acknowledgements}

\end{document}